\documentclass[12pt,a4paper]{article}
\usepackage[british]{babel}
\usepackage{epsfig}
\usepackage{amsmath}
\begin{document}
\date{}

\title{
{\vspace{-20mm} \normalsize
\hfill \parbox[t]{50mm}{\small DESY 07-078     \\
                               MS-TP-07-14     \\
                               BNL-HET-07/9 }} \\[25mm]
 Hadron masses in QCD with one quark flavour \\[5mm]}
\author{F.\ Farchioni$^{a}$,
        I.\ Montvay$^{b}$,
        G.\ M\"unster$^{a}$,\\
        E.E.\ Scholz$^{c}$,
        T.\ Sudmann$^{a}$,
        J.\ Wuilloud$^{a}$\\[5mm]
 {\small $^a$ Universit\"at M\"unster,
  Institut f\"ur Theoretische Physik,}\\
 {\small Wilhelm-Klemm-Strasse 9, D-48149 M\"unster,
  Germany}\\
 {\small $^b$ Deutsches Elektronen-Synchrotron DESY, Notkestr.\,85,
  D-22603 Hamburg, Germany}\\
 {\small $^c$ Physics Department, Brookhaven National Laboratory,
  Upton, NY 11973 USA}\\[5mm]}
%
\newcommand{\be}{\begin{equation}}
\newcommand{\ee}{\end{equation}}
\newcommand{\bea}{\begin{eqnarray}}
\newcommand{\eea}{\end{eqnarray}}
\newcommand{\half}{\frac{1}{2}}
\newcommand{\rar}{\rightarrow}
\newcommand{\lar}{\leftarrow}
\newcommand{\LCB}{\raisebox{-0.3ex}{\mbox{\LARGE$\left\{\right.$}}}
\newcommand{\RCB}{\raisebox{-0.3ex}{\mbox{\LARGE$\left.\right\}$}}}
\newcommand{\LSB}{\raisebox{-0.3ex}{\mbox{\LARGE$\left[\right.$}}}
\newcommand{\RSB}{\raisebox{-0.3ex}{\mbox{\LARGE$\left.\right]$}}}
\newcommand{\tr}{{\rm Tr}}
\newcommand{\I}{\ensuremath{\mathrm{i}\,}}
\newcommand{\E}{\ensuremath{\mathrm{e}\,}}

\maketitle

\abstract{One-flavour QCD -- a gauge theory with ${\rm SU}(3)$ colour
 gauge group and a fermion in the fundamental representation -- is
 studied by Monte Carlo simulations.
 The mass spectrum of hadronic bound states is investigated in a
 volume with extensions of $L \simeq 4.4\, r_0\; (\simeq 2.2 {\rm\, fm})$
 at two different lattice spacings:
 $a \simeq 0.37\, r_0\; (\simeq 0.19 {\rm\, fm})$ and
 $a \simeq 0.27\, r_0\; (\simeq 0.13 {\rm\, fm})$.
 The lattice action is Symanzik tree-level-improved Wilson action
 for the gauge field and (unimproved) Wilson action for the fermion.}

\newpage
\section{Introduction}\label{sec1}

 QCD with one flavour of quarks is an interesting theoretical
 laboratory to understand some aspects of the strong interaction
 dynamics, namely those not connected to spontaneous chiral symmetry
 breaking and to the existence of light pseudo-Goldstone bosons.
 As a consequence of a quantum anomaly, the U(1) axial symmetry of the
 classical Lagrangian is broken and in the limit of vanishing quark mass
 no massless Goldstone boson exists.

 An intriguing possibility at negative quark masses is the spontaneous
 breakdown of parity and charge conjugation symmetry -- a phenomenon
 first conjectured by Dashen \cite{Dashen} in the three-flavour
 theory.
 This has to do with the possible negative sign of the fermion
 determinant at negative quark masses because under the assumption
 of the positiveness of the fermion determinant Vafa and Witten
 \cite{VafaWitten} proved the impossibility of this kind of spontaneous
 symmetry breaking.

 A dramatic consequence of the absence of (broken) chiral symmetry is
 the difficulty to find a unique definition of the point with zero quark
 mass in parameter space \cite{Creutz}.
 (For an excellent summary and discussion of this problem see
 \cite{CreutzReview}.)

 Another line of recent theoretical developments is the relation
 between one-flavour ($N_f=1$) QCD and supersymmetric Yang-Mills (SYM)
 theory with one supersymmetry charge (${\cal N}=1$)
 \cite{ArmoniShifmanVeneziano}.
 This connection is the consequence of {\em orientifold planar
 equivalence} in the limit of large number of colours ($N_c \to\infty$).
 This might imply approximate relations among hadron masses even
 at $N_c=3$, for instance, the approximate degeneracy of scalar and
 pseudoscalar bound states of quarks \cite{K-HynesThacker} reflecting
 the properties of the Veneziano-Yankielovicz low energy effective
 action of ${\cal N}=1$ SYM \cite{VenezianoYankielovicz} in the mass
 spectrum of $N_f=1$ QCD.
 For instance, the mass ratio of the lowest pseudoscalar meson
 to scalar meson is predicted, including $1/N_c$ corrections, to
 be $(N_c-2)/N_c$ \cite{1overN}.
 Another prediction of orientifold equivalence is the size of the
 quark condensate in one-flavour QCD which has recently been
 compared with numerical simulation results in
 Ref.~\cite{DeGrandHoffmannSchaeferLiu}.

 In the present paper we start to explore the mass spectrum of
 hadronic states in one-flavour QCD by numerical Monte Carlo simulations.
 This requires reasonably large physical volumes at small quark masses
 and high statistics -- especially for determining glueball masses and
 contributions of disconnected quark diagrams.
 We apply the Wilson lattice fermion action which has recently been
 shown by several collaborations \cite{qq+q,CERN,QCDSF,ETMC} to be well
 suited for such an investigation.
 We start our exploratory studies here on $12^3 \cdot 24$ and
 $16^3 \cdot 32$ lattices with lattice spacing $a \simeq 0.19\,{\rm fm}$
 and $a \simeq 0.13\,{\rm fm}$, respectively.
 This means that our present setup roughly corresponds to the earlier
 simulations of the qq+q Collaboration \cite{qq+q}, but we hope to
 continue these investigations in the near future closer to the continuum
 limit as in Refs.~\cite{CERN,QCDSF,ETMC}.

 For setting the scale we use the Sommer parameter \cite{Sommer}
 $r_0$ which we set by definition to be $r_0 \equiv 0.5\,{\rm fm}$.
 In other words, whenever we speak about ``$1\,{\rm fm}$'' we always
 mean ``$2\, r_0$'' -- having in mind that one-flavour QCD is a theory
 different from QCD realised in nature.

 Since the sign of the quark determinant is a sensitive issue, we
 carefully determine it and take it into account in determining the
 expectation values.
 In the present paper we choose the quark mass to be sufficiently far
 away from zero on the positive side, where the effect of the
 determinant sign is not very strong.
 In spite of this, as we shall see, we can investigate quite small
 quark masses down to $m_q \simeq 12{\rm\,MeV}$ (that is
 $m_q r_0 \simeq 0.03$), corresponding to a pion mass
 $m_\pi \simeq 270{\rm\,MeV}$.

 Let us mention that keeping the quarks sufficiently heavy (choosing
 the hopping parameter $\kappa$ in the Wilson fermion action
 (\ref{eq02}) below $\frac{1}{8}$) the problem of negative
 quark determinants can be avoided.
 (The thermodynamics of $N_f=1$ QCD for heavy quarks have been
 investigated under this assumption in Ref.~\cite{AlexandrouThermodyn}.)
 Our aim is, however, to reach small quark masses and therefore
 we have to deal with the possibly negative sign of the quark
 determinant.

 For interpreting our results on the mass spectrum we find it useful
 to embed the $N_f=1$ QCD theory in a {\em partially quenched}
 theory with more quark flavours.
 This embedding is particularly useful if the additional quenched
 {\em valence} quark flavours have the same mass as the dynamical
 {\em sea} quark because of the exact ${\rm SU}(N_F)$ flavour symmetry
 in the combined sea- and valence-sectors ($N_F$ denotes here the total
 number of quenched and unquenched flavours).
 In most cases we consider the natural choice $N_F=3$ which is
 closest to the situation realised in nature.
 We also work out some of the predictions of {\em partially quenched
 chiral perturbation theory} (PQChPT) and compare them to the
 numerical data.

 The plan of this paper is as follows: in the next section we define
 the lattice action and briefly discuss the updating algorithm.
 In Section \ref{sec3} the partially quenched viewpoint is introduced
 and PQChPT is considered for it.
 Section \ref{sec4} is devoted to the presentation of our numerical
 simulation data.
 The last section contains a discussion and summary.

 \section{Lattice action and simulation algorithm}\label{sec2}
\subsection{Lattice action}\label{sec2.1}

 For the SU(3) Yang-Mills gauge field we apply, following
 Ref.~\cite{ETMC}, the {\em tree-level improved Symanzik} (tlSym) action
 which is a generalisation of the Wilson plaquette gauge action.
 It belongs to a one-parameter family of actions obtained by
 renormalisation group considerations and in the Symanzik improvement
 scheme \cite{Symanzik}.
 Those actions also include, besides the usual $(1\times 1)$ Wilson loop
 plaquette term, planar rectangular $(1\times 2)$ Wilson loops:
\be\label{eq01}
S_g = \beta\sum_{x}\left(c_{0}\sum_{\mu<\nu;\,\mu,\nu=1}^4
\left\{1-\frac{1}{3}\,{\rm Re\,} U_{x\mu\nu}^{1\times 1}\right\}
+c_{1}\sum_{\mu\ne\nu;\,\mu,\nu=1}^4
\left\{1-\frac{1}{3}\,{\rm Re\,} U_{x\mu\nu}^{1\times 2}\right\}
\right) \ ,
\ee
 with the normalisation condition $c_{0}=1-8c_{1}$.
 For the tlSym action we have $c_1=-1/12$ \cite{WeiszWohlert}.

 The fermionic part of the lattice action is the simple (unimproved)
 Wilson action:
\be  \label{eq02}
S_f = \sum_x \LCB \overline{\psi}_x^a\psi_x^a
-\kappa \sum_{\mu=1}^4 \left[
\overline{\psi}_{x+\hat{\mu}}^a U_{ab,x\mu}(1+\gamma_\mu)\psi_x^b
+\overline{\psi}_x^a U_{ab,x\mu}^\dagger (1-\gamma_\mu)
\psi_{x+\hat{\mu}}^b \right] \RCB \ .
\ee
 Here $\kappa$ is the hopping parameter related to the bare quark mass
 in lattice units $am_0$ by
\be\label{eq03}
\frac{1}{2\kappa} = am_0+4 \ .
\ee
 The Wilson parameter removing the fermion doublers in the continuum
 limit is fixed in (\ref{eq02})--(\ref{eq03}) to $r=1$.

\subsection{Simulation algorithm}\label{sec2.2}

 For preparing the sequences of gauge configurations a {\em Polynomial
 Hybrid Monte Carlo} (PHMC) updating algorithm was used, which is
 well-suited for theories with an odd number of fermion species.
 This algorithm is based on multi-step (actually two-step) polynomial
 approximations of the inverse fermion matrix with stochastic
 correction in the update chain as described in
 Ref.~\cite{MontvayScholz}.
 The starting point is the PHMC algorithm as introduced in
 Ref.~\cite{FrezzottiJansen,ForcrandTakaishi}.
 The polynomial approximation scheme and the stochastic correction
 in the update chain are taken over from the two-step multi-boson
 algorithm of Ref.~\cite{Montvay:tsmb}.
 For details of the updating algorithm and for notations related to
 it see Ref.~\cite{MontvayScholz}.

 In order to speed up the updating {\em even-odd preconditioning} was
 used which pushes the small eigenvalues of the (squared Hermitean)
 fermion matrix $Q[U]^2$ to larger values.
 The eigenvalues of $Q[U]^2$ are assumed to be covered on typical
 gauge configurations by the approximation interval $[\epsilon,\lambda]$.
 In exceptional cases some of the eigenvalues (typically just the
 smallest one) are outside this interval.
 In order to correct for this a {\em correction factor} $C[U]$ is
 associated with such configurations.
 The exact value of this correction factor can be written as
\be\label{eq04}
C[U] = \LCB \prod_i \left[ \lambda_i^{1/(2n_B)}\, P_1(\lambda_i)
P_2(\lambda_i) \right] \RCB^{n_B} \ .
\ee
 Here the product runs over the eigenvalues of $Q[U]^2$,
 the polynomial $P_1(x)$ is an approximation for $x^{-1/(2n_B)}$,
 $P_2(x)$ for $[x^{1/(2n_B)} P_1(x)]^{-1}$.
 The positive integer $n_B$ defines the {\em determinant break-up}
 which means that in the path integral the fermions are
 represented by
\be\label{eq05}
\left[\left( \det Q[U]^2 \right)^{1/(2n_B)} \right]^{n_B} \ .
\ee

 The part of the product in (\ref{eq04}) where $\lambda_i$ is inside
 the interval $[\epsilon,\lambda]$ can be effectively replaced
 by a stochastic estimator and then
\be\label{eq06}
C[U] = \LCB {\prod_j}^\prime \left[ \lambda_j^{1/(2n_B)}\, P_1(\lambda_j)
P_2(\lambda_j) \right] \cdot
\frac{1}{N^\prime}\sum_{n=1}^{N^\prime}
\exp{\{\eta_n^\dagger[1-P^\prime(Q[U]^2)]\eta_n\}}
\RCB^{n_B} \ .
\ee
 Here the $\prod_j^\prime$ runs over the eigenvalues outside
 the interval $[\epsilon,\lambda]$, $P^\prime(x)$ is a sufficiently
 good approximation of $[x^{1/(2n_B)} P_1(x) P_2(x)]^{-1}$,
 $N^\prime$ is the arbitrary number of stochastic estimators and
 the $\eta_n$'s are Gaussian vectors in the subspace orthogonal to
 the eigenvectors corresponding to the eigenvalues $\lambda_j$.
 In practice, one can choose the polynomial $P_2(x)$ to be such
 a good approximation that the stochastic part in (\ref{eq06})
 has no noticeable effect on the expectation values and therefore
 can completely be neglected.
 In this case the correction factor is simply given by
\be\label{eq07}
C[U] = \LCB {\prod_j}^\prime \left[ \lambda_j^{1/(2n_B)}\, P_1(\lambda_j)
P_2(\lambda_j) \right] \RCB^{n_B} \ .
\ee

 Besides the correction factor $C[U]$, the sign $\sigma[U]$ of the
 fermion determinant $\det Q[U]$ has also to be included in the
 {\em reweighting} of the configurations and then the expectation
 value of a quantity $A$ is given by
\be\label{eq08}
\langle A \rangle = \frac{\int d[U]\,\sigma[U]\,C[U]\,A[U]}
                         {\int d[U]\,\sigma[U]\,C[U]}  \ .
\ee
 This formula shows the dangerous {\em sign problem} which can arise
 due to the fluctuation of the determinant sign because in case of
 strong fluctuations of $\sigma[U]$ both nominator and denominator
 on the right hand side may become small, spoiling the statistical
 accuracy.
 (Similarly, one can also loose statistics if the correction factors
 $C[U]$ are much smaller than 1 on many configurations.)

 Typical values of the approximation interval and of the polynomial orders
 at the lightest quark mass simulated on $12^3 \cdot 24$ and
 $16^3 \cdot 32$ lattices, respectively, are collected in Table \ref{tabphmc}.
 As in Ref.~\cite{MontvayScholz}, the orders of the polynomials
 $P_j,\;(j=1,2)$ are denoted by $n_j$ and those of $\bar{P}_j,\;(j=1,2)$
 by $\bar{n}_j$, respectively.
 The simulations have been done with determinant break-up $n_B=2$. 
 (The polynomials $\bar{P}_j$ are approximating $(P_j)^{-\half}$.
 For more details see \cite{MontvayScholz} and references therein.)

 The last four columns of Table \ref{tabphmc} show the values of the
 {\em deviation norm} $\delta$ which is minimised for a given polynomial
 order $n$ in the {\em least-square approximation} scheme we are using.
 Generically $\delta$ is defined as
\begin{equation} \label{eq09}
\delta \equiv \LCB \frac{
\int_\epsilon^\lambda dx\, w(x) \left[ f(x) - P_n(x) \right]^2}
{\int_\epsilon^\lambda dx\, w(x) f(x)^2}
\RCB^\half \ .
\end{equation}
 Here $f(x)$ is the function to be approximated and $w(x)$ is a positive
 weight function actually chosen in our case to be
 $w_1(x)=w_2(x)=x^{1/(2n_B)}$ and $\bar{w}_1(x)=\bar{w}_2(x)=1$,
 respectively.
 The values of $\delta_1$ in Table \ref{tabphmc} are such that the
 average acceptance rate of the stochastic correction at the end of
 trajectory sequences is between $80-90\%$.
 The other $\delta$ values are small enough to ensure practically
 infinite precision of the expectation values.
 For more details on the algorithmic setup in our runs see also
 Section~\ref{sec4}.

\begin{table}
\begin{center}
\parbox{0.8\linewidth}{\caption{\label{tabphmc}\em
 Algorithmic parameters in the runs with lightest quark mass on
 $12^3 \cdot 24$ (first line) and $16^3 \cdot 32$ (second line) lattice,
 respectively.
 For notations see the text and also Ref.~\protect\cite{MontvayScholz}.}}
\end{center}
\begin{center}
\renewcommand{\arraystretch}{1.2}
\begin{tabular}{*{10}{|r}|}
\hline
 \multicolumn{1}{|c|}{$\epsilon$} &
 \multicolumn{1}{|c|}{$\lambda$} &
 \multicolumn{1}{|c|}{$n_1$} &
 \multicolumn{1}{|c|}{$\bar{n}_1$} &
 \multicolumn{1}{|c|}{$n_2$} &
 \multicolumn{1}{|c|}{$\bar{n}_2$} &
 \multicolumn{1}{|c|}{$\delta_1$} &
 \multicolumn{1}{|c|}{$\bar{\delta}_1$} &
 \multicolumn{1}{|c|}{$\delta_2$} &
 \multicolumn{1}{|c|}{$\bar{\delta}_2$}
 \\
\hline\hline
 $3.25\cdot 10^{-6}$ & 2.6 & 350  & 550 & 1400 & 1600 &
 $4.9\cdot 10^{-4}$ & $6.7\cdot 10^{-7}$ & $9.9\cdot 10^{-7}$ &
 $8.8\cdot 10^{-7}$ \\
\hline
 $1.2\cdot 10^{-5}$  & 2.4 & 250  & 370 & 1000 & 1150 &
 $5.4\cdot 10^{-4}$ & $8.2\cdot 10^{-7}$ & $4.8\cdot 10^{-7}$ &
 $3.1\cdot 10^{-7}$ \\
\hline
\end{tabular}
\end{center}
\end{table}

\section{Partially quenched viewpoint}\label{sec3}

 Because the classical U(1)$_A$ axial symmetry is anomalous, the
 single-flavour QCD theory does not have a continuous chiral symmetry
 apart from the U(1) quark number symmetry.
 Consequently it does not have spontaneous chiral symmetry breaking and
 hence no (pseudo-) Goldstone bosons and no easy definition of the quark
 mass \cite{Creutz}.
 In the lattice regularisation it is, however, possible to enhance the
 symmetry artificially by adding extra {\em valence quarks} which are
 {\em quenched}, that is, are not taken into account in the
 Boltzmann-weight of the gauge configurations by their fermion
 determinants.
 In principle, one might consider any number of quenched
 valence quarks with any mass values but, to remain close to QCD
 realised in nature, the most natural choice is to take two equal-mass
 valence quarks and to call them $u$ and $d$ quarks.
 The original dynamical quark can then be called $s$ quark where ``$s$''
 may stand for {\em sea} or {\em strange}.
 The theory with dynamical $s$ quark and quenched $u$ and $d$ quarks is
 {\em partially quenched}.
 (Observe that this partially quenching is somewhat unconventional,
 since some of the valence quarks are quenched but taken degenerate with
 the sea quark.)

 Using this terminology, for instance, the pseudoscalar bound state of
 $s$ and $\bar{s}$ can be called $\eta_s$.
 The corresponding scalar state is then $\sigma_s$.
 The lowest baryon state consisting of $s$ quarks, which has to have
 spin $\frac{3}{2}$ because of the Pauli principle, can be named
 $\Omega^-$ or e.g.~$\Delta_s$ etc.

 A theoretical description of partially quenched QCD can be obtained
 through the introduction of ghost quarks \cite{Morel}.
 For each (quenched) valence quark a corresponding bosonic ghost quark
 is added to the model.
 The functional integral over the ghost quark fields then cancels the
 fermion determinant of the valence quarks and only the sea quark
 determinant remains in the measure.
 In our case there are 2 flavours of valence quarks and ghost quarks,
 each, with equal masses $m_V$, and a single flavour of sea quarks with
 mass $m_S$.

 A particularly interesting point of the partially quenched theory is
 the one where all the three quark masses are equal.
 In this point there is an exact SU(3) vector-like flavour symmetry in
 the valence + sea quark sector, and the hadronic bound states appear in
 exactly degenerate SU(3)-symmetric multiplets.
 For instance, there is a degenerate octet of pseudoscalar mesons -- the
 ``pions'' ($\pi^a,\;a=1,\ldots,8$) satisfying an SU(3)-symmetric PCAC
 relation.  
 With the help of the divergence of the axialvector current $A^a_{x\mu}$
 and pseudoscalar density $P^a_x$ one can define, as usual, the bare
 {\em PCAC quark mass} $am_{\rm\scriptscriptstyle PCAC}$ in lattice
 units:
\be\label{eq10}
am_{\rm\scriptscriptstyle PCAC} \equiv
\frac{\langle \partial^\ast_\mu A^+_{x\mu}\, P^-_y \rangle}
{2\langle P^+_x\, P^-_y \rangle} \ .
\ee
 Here the indices $+$ and $-$ refer to the ``charged'' components
 corresponding to $\lambda_a \pm i\lambda_b$ (with $\lambda_{a,b}$ some
 off-diagonal Gell-Mann matrices) and $\partial^\ast_\mu$ denotes the
 backward lattice derivative.
 Due to the exact SU(3)-symmetry, the renormalised quark mass
 corresponding to $m_{\rm\scriptscriptstyle PCAC}$ can be defined by an
 SU(3)-symmetric multiplicative renormalisation:
\be\label{eq11}
m^{\rm\scriptscriptstyle R}_{\rm\scriptscriptstyle PCAC} =
\frac{Z_A}{Z_P} m_{\rm\scriptscriptstyle PCAC} \ .
\ee

 By tuning the bare quark mass on the lattice suitably, the masses of
 the ``pions'' can be made to vanish, as the numerical results
 indicate, and the renormalised quark mass vanishes, too.
 At this point the partially quenched theory has a graded
 ${\rm SU}(N_F |N_V)_L\,\otimes\,{\rm SU}(N_F | N_V)_R$ symmetry, which
 is broken spontaneously to a ``flavour'' ${\rm SU}(N_F | N_V)$.
 (Here $N_V$ is the number of additional valence quark flavours and
 $N_F \equiv N_V+N_f=N_V+1$.)
 In our case, with $N_V = 2$ flavours of valence quarks, the symmetry
 is thus SU(3$|$2).
 The ``pions'' are the Goldstone bosons of the broken SU(3) subgroup.

 Adding generic quark masses $m_V$ and $m_S$, the symmetry group is
 explicitly broken down to SU(2$|$2).
 In the special case $m_V = m_S$, considered here, the symmetry is
 still SU(3$|$2), and its subgroup SU(3) is the flavour symmetry
 mentioned above.

 The ``pions'' are, of course, not physical particles in the spectrum
 of $N_f = 1$ QCD.
 Nevertheless, their properties such as masses and decay constants
 are well defined quantities which can be computed on the lattice.
 The same is true of the PCAC quark mass
 $m^{\rm\scriptscriptstyle R}_{\rm\scriptscriptstyle PCAC}$, which is
 therefore a potential candidate for a definition of a quark mass of
 this theory.

 The relation between the pion masses and the quark masses can be
 considered in partially quenched chiral perturbation theory
 \cite{BG,SharpePQ}, including effects of the lattice spacing $a$
 \cite{ShaSi,LeeSha,Rupak-Shoresh,Aoki,BRS}.
 The pseudo-Goldstone fields are parameterized by a graded matrix
\begin{equation}
U(x) = \exp \left(\frac{\I}{F_0} \Phi(x) \right)
\end{equation}
 in the supergroup SU(3$|$2).
 (Here the normalization of $F_0$ is such that its phenomenological
 value is $\simeq 86{\rm\,MeV}$.)
 The commuting elements of the graded matrix $\Phi$ represent the
 pseudo-Goldstone bosons made from a quark and an anti-quark with equal
 statistics, and the anticommuting elements of $\Phi$ represent
 pseudo-Goldstone fermions which are built from one fermionic quark and
 one bosonic quark.
 The supertrace of $\Phi$ has to vanish, which can be implemented by a
 suitable choice of generators \cite{Sharpe-Shoresh}.

 We have calculated the masses of pseudo-Goldstone bosons in
 next-to-leading order of partially quenched chiral perturbation theory
 along the lines of Ref.~\cite{Sharpe-Shoresh}, including O($a$) lattice
 effects \cite{Rupak-Shoresh}.
 The quark masses enter the expressions in the combinations
\begin{equation}
\chi_V = 2 B_0\, m_V,\quad \chi_S = 2 B_0\, m_S,
\end{equation}
 with the usual low-energy constant $B_0$, and the lattice spacing
 occurs as
\begin{equation}
\rho = 2 W_0\, a,
\end{equation}
 where $W_0$ is another, lattice-specific, low-energy constant.
 For the pion masses we obtain
\begin{eqnarray}
m_{VV}^2 \equiv m_{\pi}^2
&=&\chi_V + \rho
     + \frac{\chi_V + \rho}{16\pi^2F_0^2}
       \left[(2 \chi_V - \chi_S + \rho)
             \ln\left(\frac{\chi_V + \rho}{16\pi^2F_0^2}\right)
             + \chi_V - \chi_S \right] \nonumber\\
&&+ \frac{8}{F_0^2} \left[
    (2L_8-L_5)\chi_V^2 + (2L_6-L_4)\chi_V\chi_S\right. \nonumber\\
&&+ \left.
    (2W_8+W_6-W_5-W_4-L_5)\rho\,\chi_V + (W_6-L_4)\rho\,\chi_S\right],
\end{eqnarray}
 where the usual low-energy parameters $L_i$ appear, together with
 addtional ones ($W_i$) describing lattice artifacts.

 The mixed mesons, whose masses $m_{VS}$ we have also calculated,
 become degenerate with the pions in the special case $m_V = m_S$.
 In this case the expression reduces to
\begin{eqnarray}
\label{eqchpt}
m_{\pi}^2
&=&\chi + \rho
    + \frac{(\chi + \rho)^2}{16\pi^2F_0^2}
      \ln\left(\frac{\chi + \rho}{16\pi^2F_0^2}\right) \nonumber\\
&&+ \frac{8}{F_0^2} \left[
        (2L_8-L_5+2L_6-L_4)\chi^2 \right. \nonumber\\
&&   \left.  +(2W_8+2W_6-W_5-W_4-L_5-L_4)\chi\rho \right].
\end{eqnarray}
 To leading order the PCAC quark mass obeys $2 B_0\,
 m^{\rm\scriptscriptstyle R}_{\rm\scriptscriptstyle PCAC} = \chi + \rho$,
 and we recognize the Gell-Mann-Oakes-Renner relation
\begin{equation}
m_{\pi}^2 = 2 B_0\, m^{\rm\scriptscriptstyle R}_{\rm\scriptscriptstyle PCAC} 
+ {\rm NLO}.
\end{equation}
 Including terms in next-to-leading (NLO) order, we can express
 $m_{\pi}^2$ in terms of
 $m^{\rm\scriptscriptstyle R}_{\rm\scriptscriptstyle PCAC}$ as
\begin{eqnarray}
m_{\pi}^2 &=& \chi_{\rm\scriptscriptstyle PCAC}
+ \frac{\chi_{\rm\scriptscriptstyle PCAC}^2}{16 \pi^2 F_0^2}
\ln \frac{\chi_{\rm\scriptscriptstyle PCAC}}{\Lambda^2} \nonumber\\
&&+ \frac{8}{F_0^2}
\left[ (2L_8-L_5+2L_6-L_4) \chi_{\rm\scriptscriptstyle PCAC}^2
\right. \nonumber\\
&&\left. + (W_8+W_6-W_5-W_4-2L_8+L_5-2L_6+L_4)
\chi_{\rm\scriptscriptstyle PCAC} \rho \right],
\end{eqnarray}
 where we define
\begin{equation}
\chi_{\rm\scriptscriptstyle PCAC}
= 2 B_0\, m^{\rm\scriptscriptstyle R}_{\rm\scriptscriptstyle PCAC} \ .
\end{equation}

 As a remark, in the case $m_V = m_S$ the masses can alternatively be
 obtained from the partially quenched theory with symmetry SU(2$|$1) by
 considering mixed pions made from a valence quark and a degenerate sea
 quark.
 Indeed, calculating the masses in this model reproduces (\ref{eqchpt}).

 The $\eta_s$ can be included in the analysis by relaxing the constraint
 of a vanishing supertrace \cite{BG,Sharpe-Shoresh}, and associating it
 with the field
\begin{equation}
\Phi_0 (x) = \mbox{s\kern-.08em Tr}\, \Phi(x).
\end{equation}
 The effective Lagrangian then contains additional terms
 depending on $\Phi_0$:
\begin{equation}
\Delta \mathcal{L} =
\alpha \partial_{\mu} \Phi_0 \partial_{\mu} \Phi_0 + m_{\Phi}^2 \Phi_0^2
+ \mathcal{O}(\Phi_0^3)\,,
\end{equation}
 where $\alpha$ and $m_{\Phi}$ are free parameters in this context.
 We content ourselves with displaying only the leading order expression
 for the mass of the $\eta_s$, which reads
\begin{equation}
m_{\eta_s}^2 =
\frac{m_{\Phi}^2 + \chi_{\rm\scriptscriptstyle PCAC}}{1 + \alpha}.
\end{equation}
 Our numerical results for $m_{\eta_s}$ allow to determine $\alpha$
 and $m_{\Phi}$.

\section{Numerical simulations}\label{sec4}

 After some preparatory search in the parameter space we concentrated
 our runs on the $12^3\cdot 24$ lattice to $\beta=3.8$ and those
 on $16^3\cdot 32$ to $\beta=4.0$.
 The parameter values, the number of analysed configurations, the average
 plaquette, its integrated autocorrelation and the value of the Sommer
 scale parameter in lattice units $r_0/a$ are summarised in
 Table~\ref{tabruns}.
 As one can see, taking the values of $r_0/a$ at highest $\kappa$'s
 (smallest quark masses), the extensions of the $12^3$ and $16^3$
 lattices are $L = 4.46\, r_0 = 2.23{\rm\, fm}$ and
 $L = 4.29\, r_0 = 2.14{\rm\, fm}$, respectively.
 Since we fix $r_0=0.5{\rm\, fm}$ by definition, these correspond to
 lattice spacings $a = 0.186{\rm\, fm}$ and $a = 0.134{\rm\, fm}$,
 respectively.

 In the update-chain by the PHMC algorithm with stochastic correction
 \cite{MontvayScholz} a {\em sequence of PHMC trajectories} is followed
 by a Metropolis accept-reject step with a higher precision polynomial.
 The total length of the trajectory sequence in the runs in
 Table~\ref{tabruns} was between 1.5 and 1.8.
 The sequences consisted out of 3-6 individual trajectories.
 The precision of the first step of polynomial approximations was
 tuned such that the acceptance of the PHMC trajectories was about
 0.80-0.85.
 The total length of the trajectory sequence was chosen such that
 the acceptance of the Metropolis test was again 0.80-0.85.
 This ensured a relatively high total acceptance of 0.64-0.72.
 During the runs we tried to optimise the parameters of PHMC.
 The different values of the {\em integrated autocorrelation times}
 for the average plaquette in Table~\ref{tabruns} are, in fact, mainly
 due to increasingly better optimisations and not so much to the
 dependence on run parameters.

 The second step approximations were more than good enough to ensure
 that the expectation values were completely unaffected by the
 remaining small imprecision.
 (See, for instance, the small relative deviations in
 Table~\ref{tabphmc}.)
 This has also been explicitly checked by performing a final stochastic
 correction on a large sample of configurations with polynomials
 $P^\prime$ of order 2500 in the stochastic part of the right hand
 side of (\ref{eq06}).

 For the calculation of the expectation values the reweighting
 procedure according to (\ref{eq08}) has to be carried out.
 For this, besides the correction factor $C[U]$ from (\ref{eq07}), also
 the sign of the fermion determinant $\sigma[U]$ is needed.
 This we calculated by the {\em spectral flow} method \cite{EdwardsFlow}.
 For the $\kappa$-dependent computation of the low-lying eigenvalues of
 the hermitean fermion matrix $Q[U]$ we followed
 Ref.~\cite{KalkreuterSimma}.

 It turned out that the effect of the correction factors
 $\sigma[U] C[U]$ is in most cases negligible.
 For instance, in run $b$ of Table~\ref{tabruns} the average value
 of $\sigma[U] C[U]$ in the denominator is 0.9982.
 In run $c$ it is 0.9842.
 In run $b$ there are 34 configurations out of 3403 where some
 eigenvalue is outside the approximation interval $[\epsilon,\lambda]$
 and out of them there is a single one with negative fermion
 determinant.
 In run $c$ there are 167 from 2884 outside $[\epsilon,\lambda]$ and
 out of them there are 26 with negative correction factor due to
 $\sigma=-1$.

 Since the sign of the fermion determinant was not determined on every
 configuration, the question arises whether perhaps some negative signs
 were missed.
 This is very improbable because we determined the sign also on the
 neighbouring configurations in addition to those with small
 eigenvalues and out of the remaining configurations we have chosen 100
 randomly for sign determination.
 None of these additional configurations turned out to have a negative
 determinant.

 In the average plaquette and $r_0/a$ the effect of the correction
 factors is completely negligible.
 For instance, in runs $b$ and $c$ the correction has an effect in the
 average value of $r_0/a$ only in the fifth digit -- whereas the
 statistical error is in the third digit.
 In all other runs besides $b$ and $c$ every eigenvalue is inside the
 approximation interval $[\epsilon,\lambda]$ and therefore, according to
 (\ref{eq07}), the correction factor is equal to 1 on every
 configuration.

\begin{table}
\begin{center}
\parbox{0.8\linewidth}{\caption{\label{tabruns}{\em
 Summary of the runs: $12^3\cdot 24$ and $16^3\cdot 32$ lattices have
 lowercase and uppercase labels, respectively.
 The number of gauge configurations, which were saved after every
 trajectory sequence, is} $N_{\textrm{conf}}$.
 {\em The average plaquette value, its autocorrelation in number of
 trajectory sequences} $\tau_{\textrm{plaq}}$ {\em and the value of
 $r_0/a$ are also given.}}}
\end{center}
\begin{center}
\renewcommand{\arraystretch}{1.2}
\begin{tabular}{*{7}{|c}|}
\hline
 \multicolumn{1}{|c|}{label} &
 \multicolumn{1}{|c|}{$\beta$} &
 \multicolumn{1}{|c|}{$\kappa$} &
 \multicolumn{1}{|c|}{$N_{\textrm{conf}}$} &
 \multicolumn{1}{|c|}{${\textrm{plaquette}}$} &
 \multicolumn{1}{|c|}{$\tau_{\textrm{plaq}}$} &
 \multicolumn{1}{|c|}{$r_0/a$}
 \\
\hline\hline
 $a$ & 3.80 & 0.1700 & 5424 & 0.546041(66) & 12.5 & 2.66(4)
 \\ \hline
 $b$ & 3.80 & 0.1705 & 3403 & 0.546881(46) & 4.6  & 2.67(5)
 \\ \hline
 $c$ & 3.80 & 0.1710 & 2884 & 0.547840(67) & 7.6  & 2.69(5)
 \\ \hline\hline
 $A$ & 4.00 & 0.1600 & 1201 & 0.581427(36) & 4.3  & 3.56(5)
 \\ \hline
 $B$ & 4.00 & 0.1610 & 1035 & 0.582273(36) & 4.1  & 3.61(5)
 \\ \hline
 $C$ & 4.00 & 0.1615 & 1005 & 0.582781(32) & 3.3  & 3.73(5)
 \\ \hline\hline
\end{tabular}
\end{center}
\end{table}

\subsection{Results for hadron masses}\label{sec4.1}

\begin{table}[t]
\begin{center}
\parbox{0.8\linewidth}{\caption{\label{tab:hadmass}\em
 Results for light hadron masses in $N_f=1$ QCD.}}
\end{center}
\begin{center}
\renewcommand{\arraystretch}{1.2}
\begin{tabular}{*{6}{|c}|}
\hline
 \multicolumn{1}{|c|}{run} &
 \multicolumn{1}{|c|}{$am_{\eta_s}$} &
 \multicolumn{1}{|c|}{$am_{\sigma_s}$} &
 \multicolumn{1}{|c|}{$am_{0^{++}}$} &
 \multicolumn{1}{|c|}{$am_{\Delta_s}$} 
 \\
\hline\hline
 $a$ & 0.462(13)     & 0.660(39)       & 0.777(11)     & 1.215(20)
 \\ \hline
 $b$ & 0.403(11)     & 0.629(29)       & 0.685(10)     & 1.116(38)
 \\ \hline
 $c$ & 0.398(28)     & 0.584(55)       & 0.842(16)     & 1.204(57)
 \\ \hline\hline
 $A$ & 0.455(17)     & 0.607(57)       & 1.083(79)     & 1.006(15)
 \\ \hline
 $B$ & 0.380(18)     & 0.554(52)       & 1.032(66)     & 0.960(15)
 \\ \hline
 $C$ & 0.316(22)     & 0.613(67)       & 0.980(97)     & 0.876(26)
 \\ \hline\hline
\end{tabular}
\end{center}
\end{table}

 Starting with the mesonic states, we consider the simplest
 interpolating operators in the pseudoscalar and scalar sectors:
\bea
0^+:   \quad P(x)&=& \bar\psi(x)\gamma_5\psi(x)~, \\
0^-: \quad S(x)&=& \bar\psi(x)\psi(x)~.
\eea
 We denote with $\eta_s$ and $\sigma_s$ the corresponding hadron states
 at the lowest end of the energy spectrum (the usual notation $J^P$
 is used for the respective quantum numbers).
 Corresponding states in the QCD spectrum with the same quantum numbers
 are the $\eta^\prime(958)$ and $f_0(600)$ (or $\sigma$).
 (Note, however, that the states in QCD are linear combinations of
 $\bar{u}u$, $\bar{d}d$ and $\bar{s}s$ components -- in contrast to
 the states in $N_f=1$ QCD which are built out of a single quark
 flavour.)

 In the case of the pseudoscalar mesons, invariance under the flavour
 group plays a special role when comparing with QCD states  because of
 the  U(1) axial anomaly.
 (This is not the case for baryons, see the following.)

 Analogously to flavour singlet mesons in QCD, the correlators of the
 above interpolating operators contain disconnected diagrams.
 These were computed by applying stochastic estimator techniques (SET),
 and in particular the variant of~\cite{set} with $Z_2$ noise and spin
 dilution.
 The method was already applied to the case of SYM~\cite{FaPe} (as
 mentioned in the introduction, SYM shares many similarities with
 $N_f=1$ QCD).
 In order to optimize the computational load, taking also
 autocorrelations into account, every fifth configuration was typically
 analysed, with 20 stochastic estimates each.

 Spin~0 states can be also build by purely gluonic operators.
 These are a well known object of investigation in lattice QCD
 were they should describe the glueballs.
 Due to the expected signal-noise ratio of their purely gluonic
 correlation they belong to the most notorious particles to measure.
 In particular the $0^{++}$ glueball has the same quantum numbers as
 the $\sigma_s$ meson.
 As a consequence, these two states can also mix with each other
 but in this first investigation we neglect the mixing and consider
 only diagonal correlators for both states.

 We used the single spatial plaquette to obtain the mass of the
 $0^{++}$ ground state.
 To increase the overlap of the operator with this state we used APE
 smearing and also performed variational methods to obtain optimal
 glueball operators from linear combinations of the basic operators.

 We now come to the baryon sector.
 The simplest baryonic interpolating field which can be built 
 out of  one quark flavour is
\be
 {\Delta_i}(x)\:=\:\epsilon_{abc}[\psi_a(x)^TC\gamma_i\psi_b(x)]\psi_c(x)\ .
\ee
 The above operator also contains a spin 1/2 component implying that
 the spin 3/2 component, on which we focus, must be projected out from
 the spinorial correlator
\be
G_{ji}(t)=\sum_{\vec{x}}
\left\langle \Delta_j(\vec{x},t)\bar\Delta_i(0)\right\rangle \ .
\ee
 We follow~\cite{regina} and consider the spin-projected correlator
\be
G_{3/2}(t)=\frac{1}{6}{\rm Tr}
 \left[ G_{ji}(t)\gamma_j\gamma_i+G_{ii}(t)\right]\ .
\ee
 The low lying hadron state contributing to the above correlator is
 expected to have positive parity $(\frac{3}{2}^+)$.
 This corresponds to the $\Delta(1232)^{++}$ of QCD if our dynamical
 fermion is interpreted as an $u$ quark.
 If the dynamical fermion is taken to be the $s$ quark then this would
 be the $\Omega^-$ baryon.
 (However, spin and parity of the corresponding particle have not been
 yet measured, so the identification of this state with the
 $\Omega^-$ baryon is still uncertain~\cite{pdg}).
 In correspondence to $\eta_s$ and $\sigma_s$, in what follows we call
 this state $\Delta_s$.
 (Here one can interpret the index $s$ as referring to the ``sea''
 quark.)

 It should be noted at this point that the above QCD states are not
 flavour singlets in $N_f=3$ QCD (and in the one flavor partially
 quenched theory).
 We recall here that interpolating fields corresponding to flavour
 singlet baryon states cannot be build in QCD if only quark fields are
 considered as ingredients.

\begin{figure}[ht]
  \centering
  \includegraphics[angle=-90,width=0.7\textwidth]{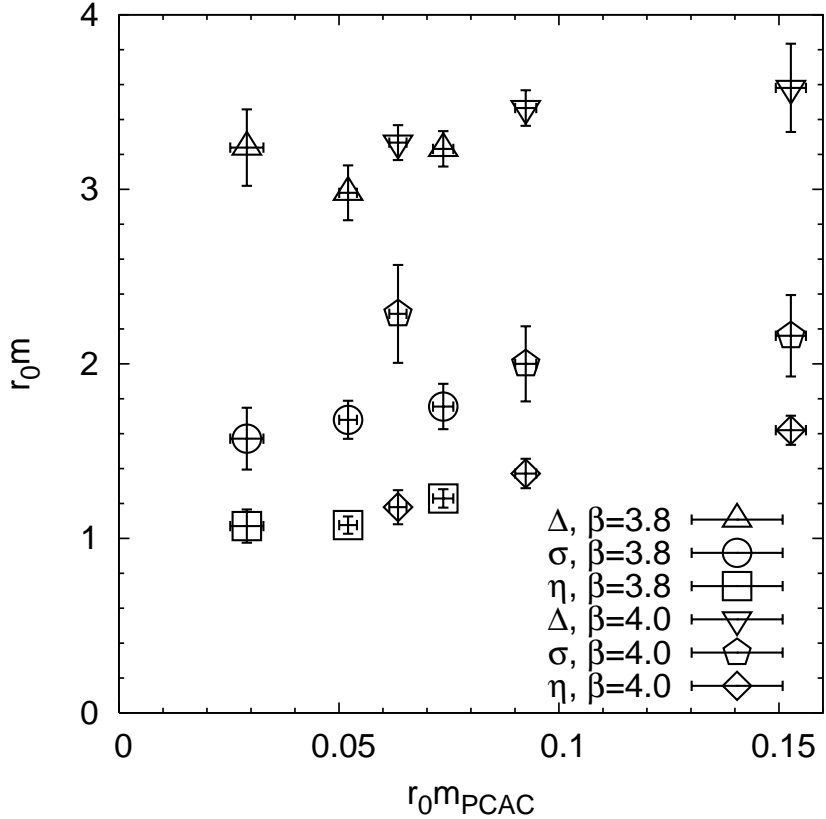}
  \parbox{0.8\linewidth}{\vspace*{-1em}\caption{\label{fig:hadmass}\em
 The mass of the lightest physical particles in one-flavour QCD as a
 function of the PCAC quark mass.
 The masses are multiplied by the scale parameter $r_0$
 in order to obtain dimensionless quantities.
 Open and full symbols refer to $\beta=3.8$ and $\beta=4.0$,
 respectively.}}
\end{figure}

 The results of the hadron masses are reported in
 Table~\ref{tab:hadmass} and, as a function of the bare PCAC quark mass
 $m_{\rm \scriptscriptstyle PCAC}$, in Fig.~\ref{fig:hadmass}.
 In the figure the masses are multiplied by the Sommer scale parameter
 $r_0$, therefore one can put the results for both lattice spacings in a
 single plot and check their scaling.
 (The expected small change of the multiplicative renormalisation
 factor of $m_{\rm\scriptscriptstyle PCAC}$ between $\beta=3.8$ and
 $\beta=4.0$ is neglected here.)

 Only in the case of run $c$ the measurement correction has a sizeable
 effect on the mass estimates.
 In this case configurations with negative determinant where singled
 out: the sign of the determinant has the effect of pushing the
 masses up by $7-10$~\%.

 The errors on the glueball mass are rather large -- especially on the
 $16^3 \cdot 32$ lattice at $\beta=4.0$ -- therefore they are not
 shown in the figure.
 Obviously, our statistics is not sufficient for this purpose.
 In general a larger number of configurations would improve the
 determinations in the glueball sector.
 Since the computational load is in this case negligible, for future
 runs we plan a more frequent storage of the gauge configuration.

\subsubsection{Valence analysis}\label{sec4.1.1}

 The connected contribution to the meson correlators can be interpreted
 as a non singlet meson made up of valence quarks in the partially
 quenched picture, see Sec.~\ref{sec3}. 
 The pseudoscalar channel corresponds in particular to the ``valence''
 pion. 
 Since the computation of the connected diagrams is less demanding, we
 could afford the analysis of the complete set of configurations.

 In the baryon sector, one can define a ``valence'' nucleon, with the
 usual projector operator
\be
N(x)=\epsilon_{abc}[\psi_a(x)^TC \psi^\prime_b(x)]\psi_c(x)\ ,
\ee
 where $\psi^\prime$ can be interpreted as the field of the valence
 quark.

\begin{figure}[ht]
  \centering
  \includegraphics[angle=-90,width=0.7\textwidth]{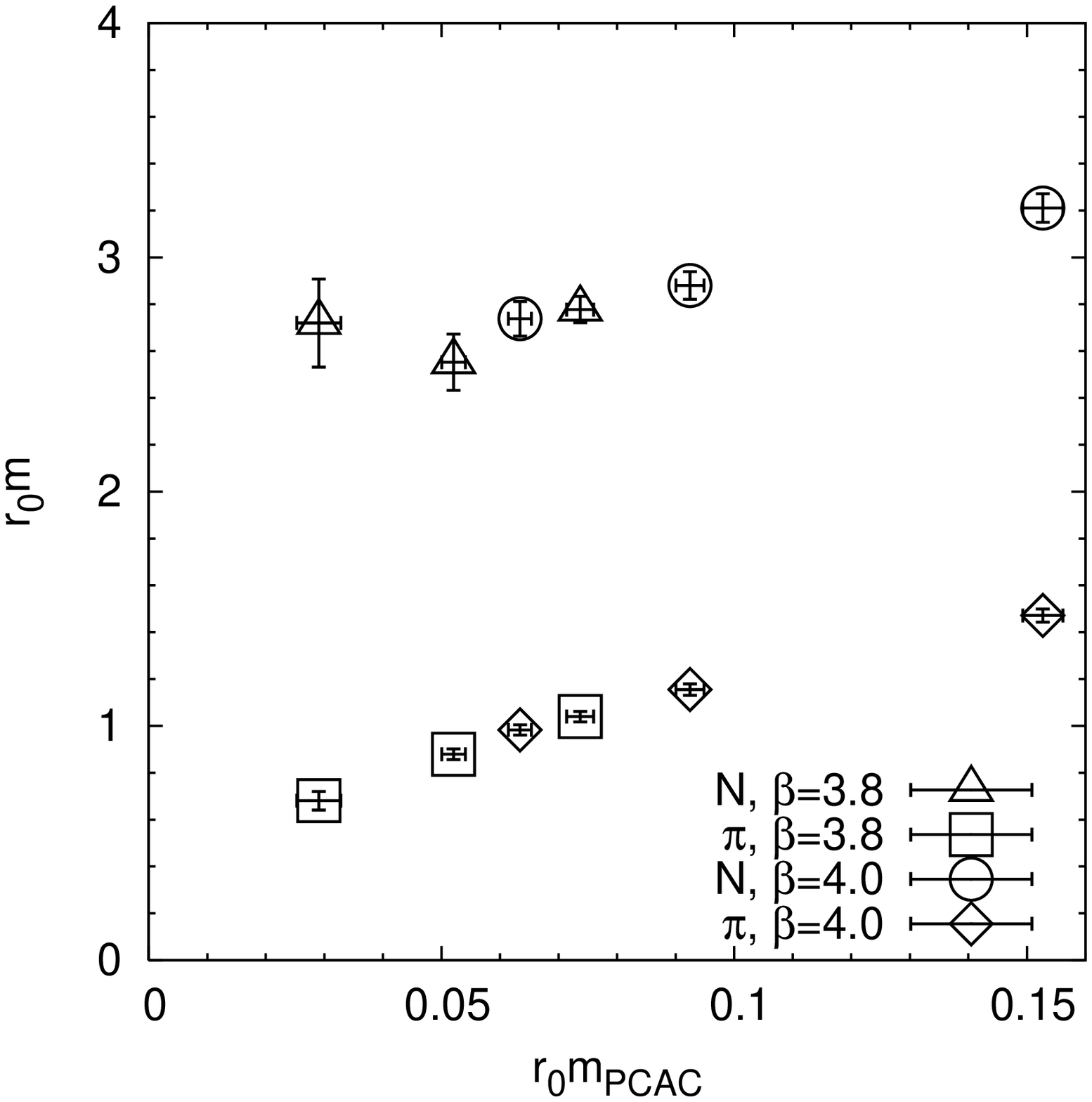}
  \parbox{0.8\linewidth}{\vspace*{-1em}\caption{\label{fig:valhad}\em
 The mass of the valence pion and nucleon as a function of the bare
 PCAC quark mass.
 Open and full symbols refer to $\beta=3.8$ and $\beta=4.0$,
 respectively.}}
\end{figure}

 The results concerning valence hadron masses are reported in
 Table~\ref{tab:valhad} and Fig.~\ref{fig:valhad}.
 In addition, the bare PCAC quark mass according to the definition in
 (\ref{eq10}) and the bare pion decay constant in lattice units $af_\pi$
 are also included.
 $f_\pi$ and its renormalised counterpart $f^{\rm\scriptscriptstyle R}_\pi$
 are defined as
\be
af_\pi = (am_\pi)^{-1} \langle 0 | A^+_{x=0,\mu=0} | \pi^-(\vec{p}=0) \rangle \ ,
\hspace{4em}
f^{\rm\scriptscriptstyle R}_\pi = Z_A f_\pi
\ee
 where $A^+_{x\mu}$ is the axialvector current as in (\ref{eq10}) and
 $\pi^-(\vec{p}=0)$ is a pion state with zero momentum.
 (The normalisation of $f_\pi$ is such that in nature we have
 $f^{\rm\scriptscriptstyle R}_\pi \simeq 130{\rm\,MeV}$.)
 The value of $af_\pi$ on the lattice is obtained by the method
 described in \cite{Lattice:fpi}.
 In Fig.~\ref{fig:valhad} the masses are multiplied by the scale
 parameter $r_0$ in order to obtain dimensionless variables.

\begin{table}
\begin{center}
\parbox{0.8\linewidth}{\caption{\label{tab:valhad}\em
 The PCAC quark mass $m_{\rm\scriptscriptstyle PCAC}$, the pion mass
 $m_\pi$ and decay constant $f_\pi$, and the nucleon mass $m_N$ in
 lattice units.}}
\end{center}
\begin{center}
\renewcommand{\arraystretch}{1.2}
\begin{tabular}{|c|*{4}{|l}|}
\hline
 \multicolumn{1}{|c|}{run} &
 \multicolumn{1}{|c|}{$am_{\rm\scriptscriptstyle PCAC}$} &
 \multicolumn{1}{|c|}{$am_\pi$} &
 \multicolumn{1}{|c|}{$af_\pi$} &
 \multicolumn{1}{|c|}{$am_N$} 
 \\
\hline\hline
 $a$ &   0.02771(45) &   0.3908(24) &  0.1838(11) &  1.0439(54)
 \\ \hline
 $b$ &   0.01951(39) &   0.3292(25) &  0.1730(15) &  0.956(27)
 \\ \hline
 $c$ &   0.0108(12)  &   0.253(10)  &  0.156(10)  &  1.011(51)
 \\ \hline\hline
 $A$ &   0.04290(36) &   0.4132(21) &  0.1449(9)  &  0.9018(44)
 \\ \hline    
 $B$ &   0.02561(31) &   0.3199(22) &  0.1289(10) &  0.7978(53)
 \\ \hline
 $C$ &   0.01700(30) &   0.2635(24) &  0.1188(12) &  0.734(10)
 \\ \hline\hline
\end{tabular}
\end{center}
\end{table}

\subsubsection{Chiral Perturbation Theory fits}\label{sec4.1.2}

\begin{figure}[ht]
  \centering
  \includegraphics[angle=-90,width=0.7\textwidth]{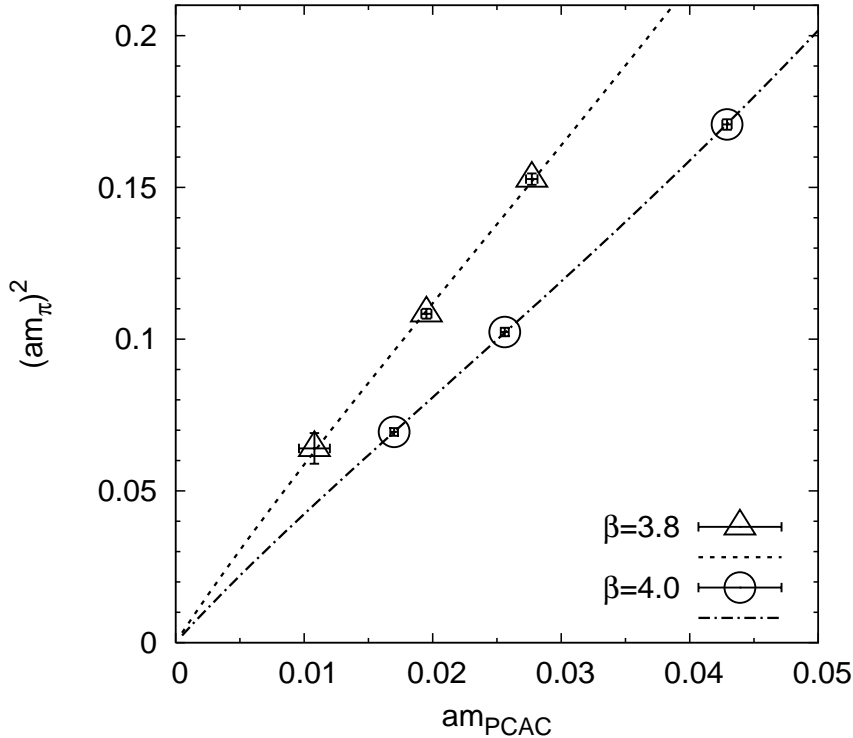}
  \parbox{0.8\linewidth}{\vspace*{-1em}\caption{\label{fig:mpi}\em
 Pion masses squared in lattice units and the results of the PQChPT
 fit.}}
\end{figure}

\begin{figure}[ht]
  \centering
  \includegraphics[angle=-90,width=0.7\textwidth]{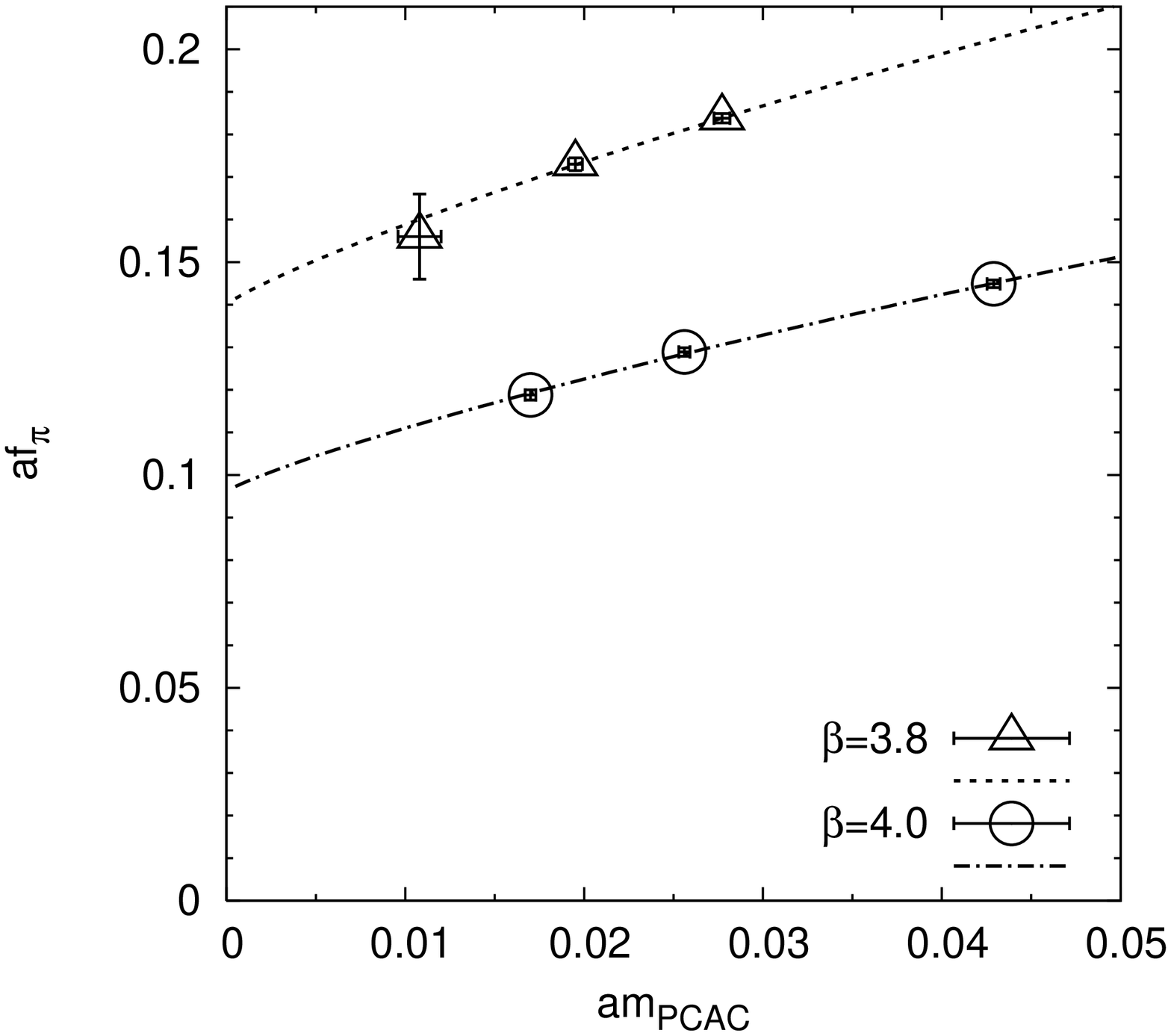}
  \parbox{0.8\linewidth}{\vspace*{-1em}\caption{\label{fig:fpi}\em
 Pion decay constants in lattice units and the results of the PQChPT
 fit.}}
\end{figure}

 The properties of the valence pion (pion mass $m_\pi$ and decay
 constant $f^{\rm\scriptscriptstyle R}_\pi$) can be analysed in
 partially quenched ChPT.
 We fit $a^2m_\pi^2$ and $af_\pi$ simultaneously as a function of
 $am_{\rm\scriptscriptstyle PCAC}$ including the data at both values of
 $\beta$.
 There are not enough data in order to account for the lattice
 artifacts.
 Therefore the fit is done with the continuum formulae
\begin{equation}
m_{\pi}^2 = \chi_{\rm\scriptscriptstyle PCAC}
+ \frac{\chi_{\rm\scriptscriptstyle PCAC}^2}{16 \pi^2 F_0^2}
\ln \frac{\chi_{\rm\scriptscriptstyle PCAC}}{\Lambda_3^2}\,,
\qquad
\frac{f^{\rm\scriptscriptstyle R}_{\pi}}{F_0 \sqrt{2}} =
1 - \frac{\chi_{\rm\scriptscriptstyle PCAC}}{32 \pi^2 F_0^2}
\ln \frac{\chi_{\rm\scriptscriptstyle PCAC}}{\Lambda_4^2}\,,
\end{equation}
 with the low-energy constants
\bea
\Lambda_3 &=& 4\pi F_0 \exp \{ 64 \pi^2 (L_4 + L_5 - 2L_6 - 2L_8) \}\,,
\nonumber \\
\Lambda_4 &=& 4\pi F_0 \exp \{ 64 \pi^2 (L_4 + L_5) \}\,.
\eea
 The changes of the renormalisation constants $Z_A$, $Z_P$ between the
 two $\beta$ values are neglected.
 The results are displayed in Figs.~\ref{fig:mpi} and \ref{fig:fpi}.


 Owing to the fact that the number of degrees of freedom in the fit is
 small, the uncertainty of the fit parameters is relatively large.
 The determination of the universal low-energy scales $\Lambda_3/F_0$
 and $\Lambda_4/F_0$ can be improved by considering the
 ratios~\cite{AlphaRatio,qq+q}
\begin{equation}
\frac{m_\pi^2}{m_{\pi,\rm ref}^2}\,,\quad
\frac{f_\pi}{f_{\pi,\rm ref}}\,,
\end{equation}
 in which some of the coefficients cancel.
 We consider the data on the larger lattice at $\beta=4.0$ and take the
 quantities at $\kappa=0.1615$ as reference.
 The fit yields
\begin{eqnarray}
\frac{\Lambda_3}{F_0}&=&10.0 \pm 2.6\,,\\
\frac{\Lambda_4}{F_0}&=&31.5 \pm 14.3\,,
\end{eqnarray}
 which is compatible with the phenomenological values from ordinary
 QCD~\cite{Duerr}.

 In order to estimate the parameters $\alpha$ and $m_\Phi$, related to
 the mass of the $\eta_s$ (see Sec.~\ref{sec3}), we made a fit of
 $m_\pi^2$ and $m_{\eta_s}^2$ at $\beta=4.0$ in leading-order ChPT.
 The result is
\begin{equation}
\alpha = -0.03(19)\,,\quad
am_\Phi = 0.18(8)\,,
\end{equation}
 indicating the vanishing of $\alpha$.
 Fixing $\alpha=0$ in the fit yields
\begin{equation}
am_\Phi = 0.19(2)\quad{\rm or}\quad
r_0m_\Phi = 0.72(10)\,,
\end{equation}
 where the value of $r_0/a$ extrapolated to vanishing PCAC quark mass is
 used.

 This constant, whose value in physical units is
 $m_{\Phi} = 284(40){\rm\,MeV}$, can be related to the quenched
 topological susceptibility $\chi_t$ through
 the Witten-Veneziano formula \cite{WittenVeneziano}
\be
m_{\Phi}^2 = \frac{4 N_f}{(f_\pi^{\rm\scriptscriptstyle R})^2} \chi_t\,,
\ee
 which is valid in leading order of the $1/N_c$ expansion.
 With $\chi_t = (193 \pm 9{\rm\,MeV})^4$ \cite{chit} and our value for
 $f_{\pi}^{R}$ we would obtain $m_{\Phi} = 426{\rm\,MeV}$.

\section{Discussion}\label{sec5}

 This first Monte Carlo investigation of the hadron masses in QCD
 with $N_f=1$ dynamical quark flavour reveals the qualitative features
 of the low lying particle spectrum in this theory.
 The spatial extensions of our $12^3\cdot 24$ and $16^3\cdot 32$
 lattices are about $L \simeq 2.2{\rm\,fm}$ (see
 Table~\ref{tabruns}).\footnote{In order to have some relation to the
 scales in real QCD, we set the Sommer scale parameter by definition to
 be $r_0\equiv 0.5{\rm\,fm}$}
 This implies lattice spacings $a \simeq 0.19{\rm\,fm}$ and
 $a \simeq 0.13{\rm\,fm}$, respectively.
 The (bare) quark masses are reasonably small -- in a range
 10-30{\rm\,MeV} and 25-60{\rm\,MeV} on the $12^3\cdot 24$ and
 $16^3\cdot 32$ lattice, respectively.
 The updating algorithm we use (PHMC with stochastic correction
 \cite{MontvayScholz}) works fine in this range making the extension
 of the Monte Carlo investigations towards larger volumes, smaller
 quark masses and smaller lattice spacings straightforward.
 In the present runs the fluctuation of the eigenvalues of the
 fermion matrix towards exceptionally small (or negative) values can be
 easily handled by reweighting the configurations during the evaluation
 of expectation values.
 In fact, except for the run with the smallest quark mass on the
 $12^3\cdot 24$ lattice where the reweighting has a small effect,
 the reweighting is completely negligible or even unnecessary.

 The lightest hadron is the pseudoscalar meson bound state of a
 quark and an antiquark -- the $\eta_s$-meson (see
 Table~\ref{tab:hadmass} and Figure~\ref{fig:hadmass}).
 The corresponding scalar bound state -- the $\sigma_s$-meson -- is in
 our points by about a factor 1.5 heavier.
 Compared to the estimate in \cite{1overN}
 $m_{\sigma_s}/m_{\eta_s} \simeq N_c/(N_c-2)=3$ this result is too low
 but the situation could be better in the zero quark mass limit where
 the prediction of \cite{1overN} applies to.
 The lightest baryon -- the $\Delta_s$-baryon -- is by a factor of about
 3 heavier than the $\eta_s$-meson.
 The lightest glueball lies between the $\sigma_s$-meson and the
 $\Delta_s$-baryon, but its mass could not be properly measured on the
 $16^3\cdot 32$ lattice with our statistics.
 In general, the mass measurements have relatively large errors --
 between 3-10\% -- and no infinite volume and continuum limit
 extrapolations could be performed with our present data.
 We hope to return to these questions and to give more precise
 results in future publications.

 An interesting aspect of $N_f=1$ QCD is the possibility of a
 {\em partially quenched} extension with valence quarks.
 In particular, adding two valence quarks, the model has similarities
 to QCD in nature with its three light ($u$, $d$ and $s$) quark
 flavours.
 A theoretically interesting special case is if all three quarks,
 the dynamical one and the two valence ones, have exactly equal masses.
 In this case there is an exact ${\rm SU}(3)$ flavour symmetry.
 This can be exploited for the introduction of a quark mass by
 defining it as the PCAC quark mass in the partially quenched theory.
 In this extended model there exist the usual light hadron states well
 known from real QCD: the pseudoscalar pseudo-Goldstone bosons
 (pions etc.), the nucleon etc.
 The results for the masses of the lightest states and the decay
 constant of the pseudoscalar bosons are collected in
 Table~\ref{tab:valhad} and also shown in Figure~\ref{fig:valhad}.

 Since the physical volumes of the $12^3$ and $16^3$ lattices are
 to a good approximation equal, the comparison of the results at
 the two different lattice spacings gives a hint for the magnitude
 of the deviations from the continuum limit.
 As one can see in Figs.~\ref{fig:hadmass} and \ref{fig:valhad},
 the scaling between $\beta=3.8$ and $\beta=4.0$ is reasonably good --
 especially for the lightest states $\eta_s$ and $\pi$.
 However, for reliable continuum limit estimates more data at several
 lattice spacings are required.

 In the pseudoscalar sector of the partially quenched model one can
 apply partially quenched Chiral Perturbation Theory for fitting the
 mass and the decay constant.
 As Figs.~\ref{fig:mpi} and \ref{fig:fpi} show, the NLO formulae
 give good fits but the number of degrees of freedom in the fits is
 small and therefore the uncertainty of the fit parameters is
 relatively large.

\vspace*{2em}
{\large\bf Acknowledgments}

\noindent
 We are grateful to Luigi Scorzato for valuable discussions and for
 helping us in the set-up of the programs for investigating the
 eigenvalue spectrum of the fermion matrix.
 We thank the computer centers at DESY Hamburg and NIC at
 Forschungszentrum J{\"u}lich for providing us the necessary technical
 help and computer resources.
 This work is supported in part by the Deutsche Forschungsgemeinschaft
 under grant Mu757/13-1.
 E.S. is supported by the U.S.\ Dept.\ of Energy under contract
 DE-AC02-98CH10886.

\vspace*{2em}


\end{document}